# A Proposed Access Control-Based Privacy Preservation Model to Share Healthcare Data in Cloud

Pankaj Khatiwada, Hari Bhusal, Ayan Chatterjee, Martin W. Gerdess
Faculty of Engineering and Science
University of Agder
Grimstad, Norway
pankaj.khatiwada@uia.no

*Abstract:* Healthcare data in cloud computing facilitates the treatment of patients efficiently by sharing information about personal health data between the healthcare providers for medical consultation. Furthermore, retaining the confidentiality of data and patients' identity is a another challenging task. This paper presents the concept of an access control-based (AC) privacy preservation model for the mutual authentication of users and data owners in the proposed digital system. The proposed model offers a high-security guarantee and high efficiency. The proposed digital system consists of four different entities, user, data owner, cloud server, and key generation center (KGC). This approach makes the system more robust and highly secure, which has been verified with multiple scenarios. Besides, the proposed model consisted of the setup phase, key generation phase, encryption phase, validation phase, access control phase, and data sharing phase. The setup phases are run by the data owner, which takes input as a security parameter and generates the system master key and security parameter. Then, in the key generation phase, the private key is generated by KGC and is stored in the cloud server. After that, the generated private key is encrypted. Then, the session key is generated by KGC and granted to the user and cloud server for storing, and then, the results are verified in the validation phase using validation messages. Finally, the data is shared with the user and decrypted at the user-end. The proposed model outperforms other methods with a maximal genuine data rate of 0.91.

1. INTRODUCTION

Cloud-enabled platforms are essential for delivering cost-effective health services with higher scalability and ubiquitous network access. Transferring Electronic Health Records (EHRs) to the cloud poses major threats to privacy and confidentiality [8]. Besides, there are several non-standardized communication architectures used in previous works that cause semantic divergences,which reveals that healthcare providers are responsible for data confidentiality with appropriate access controls[9][14].

Tampering is a challenge related to the confidentiality of personal health generated data ("PGD") during transmission through the wireless communication channels. The majority of the patients are concerned about the confidentiality of PGD in fear of unauthorized disclosure of data or illegitimate access to data [12][10]. It is commonly recognized that cloud computing, as well as open standards, are the most significant keystones for streamlining healthcare for the health records maintaining, managing the diseases, collaboration with the peers, and data analysis. Enabling access to the healthcare ubiquitous helps to access the healthcare data anytime from anywhere, with reduced cost and time. Therefore, the effective enforcement of privacy and security is one of the key criteria of healthcare success in a cloud infrastructure [11][13].

It is mandatory to guarantee privacy as well as security of the data for verifying the identity of the entity and creating the shared session key for the communication. Several key agreements and authentication schemes are presented with their security issues in literature. In [16], a brief analysis of monitoring human conditions using wireless sensors is presented where the smart health system is utilized to join the human activities of wearable devices continuously. In [18], the anonymous health records deduction approach is devised to monitor people's health conditions. The ideal IoT-enabled medical system is introduced for collecting IoT-based data and transferring it to the required users in [17]. But, this method does not focus on data security and privacy in the health system. Consequently, the researchers developed two types of attribute-driven encryption mechanisms. In [15], abuse-free attribute enabled Access Control (AC) is developed for producing data security to the cloud users and authors present a very primitive idea for dealing with the AC system. In [20], the cipher-text policy system driven access control system was introduced, which enhanced the accountability of both the communicating parties. However, this scheme supported only a selective model for security. The number of communication messages is key, also increasing the overhead of the transmitting messages. In [19], the previous medical authentication protocol is designed for predicting the suffers from the smart card stolen attack.

This study aims to design the access control-based privacy preservation method to enable secure data sharing in cloud storage. The proposed AC-based privacy preservation model is composed of four entities, such as user, data owner, cloud server, and the KGC. Moreover, the developed model is enabled based on six phases, namely setup, key generation, encryption, validation, access control, and data sharing. In the setup phase, the user is registered in the cloud server based on the message to generate the master key and system security parameter in KGC. After that, the public key is identified in KGC, and followed by stored in the cloud server and the user. Then the data is encrypted to determine cipher-text and is stored in the cloud server. Once the encryption phase is completed, the session key is created in the access control phase and is saved in the cloud server. Consequently, the validation phase is done using two validation messages, and finally, the data is shared with the user, and the data is decrypted at the user side to maintain data privacy in the confidential health care data.

The remaining sections of the paper are arranged as follows: Section 2 explains the review of the existing access control-based privacy preservation method and its challenges. Section 3 elaborates on the proposed method for AC-based privacy preserved data sharing. Section 4 explains the result and discussion of the developed approach, and finally, the conclusion is made in section 5.

## 2. MOTIVATION

This section describes the review of various access control-based privacy preservation methods along with their merits and demerits. It will motivate researchers to develop AC-based privacy preservation.

### 2.1 Literature survey

The eight classical strategies based on access control-based privacy preservation along with its limitations, as described as follows - Yang Yang *et al.* [1] developed a secure system for devising the two-fold access control approach for emergency and normal conditions. In the typical application, the healthcare staff with attribute secret keys having data access privilege. Also, the historical medical data of patients were analyzed based on the password-enabled break glass access method. For storing the storage overhead in the big data, the secure deduplication technique was introduced for eliminating duplicate medical files with the same data that leads to encrypt with various access policies. The method needs a lot of storage space and produces inflexibility to the system. Bander A. Alzahrani *et al.* [2] developed patient healthcare monitoring and the authentication protocol for Wireless Body Area Networks (WBAN). This method was utilized to perform lightweight operations with few security loopholes. In addition, an authenticated key agreement protocol was introduced for WBAN. The method gets real-time patient status and the information but failed to explore the efficiencies and optimizations in the authentication protocol.

Bhawna Narwal and Amar Kumar Mohapatra [3] presented a secure energy-efficient-based mutual authentication and the key agreement (SEEMAKA) scheme for the WBAN topology. This method achieved the properties of desirable security with several security attacks based on bitwise XOR operations and fewer hash invocations for meeting less capable sensor nodes. Hence, the security was assessed based on sound informal analysis and Automated Validation of Internet Security Protocols and Applications (AVISPA). However, SEEMAKA and Burrows–Abadi–Needham (BAN) was introduced to identify the correctness of the developed model but leads to poor processing costs. Saurabh Rana and Dheerendra Mishra [4] presented cipher-text attribute-enabled encryption (CP-ABE) for providing encrypted data security. Here, the previous access policies were in cleartext form to access the encrypted sensitive data. Also, the small attribute universe was supported for restricting practical deployments of the method. Furthermore, efficient access control was devolved and designed for medical services. The method was unable to deal with the attribute anonymity to protect data privacy.

### 2.2 Challenges

The challenges confronted by the conventional strategies are deliberated below:

- In [1], privacy-preserving smart IoT-enabled healthcare system is devised for ensuring patients' healthcare data securely. Because of security protection, the healthcare data is encrypted by various medical staff in the storage system. This cipher-text is outsourced to the public cloud that occupies the vast storage space simultaneously.

- The patient's privacy in the WBAN networks is challenging because of its mobility and openness. The adversaries attempt to intrude on the privacy of the patients and initiate the other attacks [2].

- Despite potential benefits, the method in [3] is resource scanty with limited communication bandwidth, limited battery energy/power, low computational power, and circumscribed memory space. In addition, security and privacy leakage of transmitted physiological data in the open channel with infrastructure are some of the formidable challenges [3].

- The presence of continuous data is the main focus in the healthcare industry, which is suitable for patient treatment. Moreover, the accessibility and availability of the uninterrupted data are of concern even in user error or hardware failure [7].

- In [5], the biometric, password-based authentication approach for the WHSNs is devised for attaining privacy preservation in the cloud. However, the problem with the sensor nodes is that they are equipped with less battery power. Therefore, energy efficiency and security become major problems.

## 3. THE PROPOSED ACCESS CONTROL MECHANISM FOR PRIVACY PRESERVED DATA SHARING

This section presents the proposed access control-based privacy preservation model for data sharing using cloud infrastructure. This work designs a model for the mutual authentication of user and data owner in the system. However, the system consists of four different entities, such as user, data owner, cloud server, and Key generation center (KGC). The proposed access control-based privacy preservation model involves six different phases, namely setup phase, key generation phase, encryption phase, access control phase, validation phase, and data sharing phase. The setup phase is run by the data owner, which takes input as a security parameter and generates a system master key and public parameter. The next step is the key generation phase to generate private keys for attaining security. The key is used to encrypt and decrypt data whenever data is being encrypted or decrypted. In the encryption phase, the data are encrypted using the encryption algorithm by taking the shared file as input. It generates the cipher-text, file encryption key, and keyword set from the file, which is then validated and verified in the validation phase. The risk of unauthorized access to the cloud is minimized in the access control phase, and finally, data privacy is achieved in the data privacy phase. Figure 1 illustrates the schematic view of the proposed access control-based privacy preservation model for healthcare data sharing with cloud infrastructure.

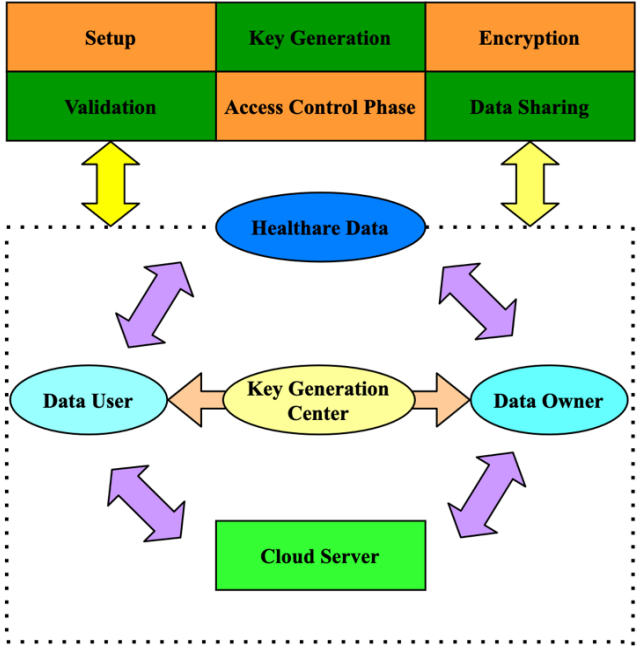

**Figure 1.** Schematic view of proposed access control approach for privacy preserved data sharing

*3.1 Setup phase*

The initial phase of the proposed access control-based privacy preserved data sharing is the setup phase. The setup phase is run by three entities, such as the user, cloud server, and the KGC. Initially, the user generates user-ID $U_{ID}$ and the password $U_{ps}$ and sends it to the cloud server. The cloud server received the user-ID and password from the user and saved them in the cloud server as $U_{ID}^*$ and $U_{ps}^*$. After that, the cloud server generates the message using the $U_{ID}$ and system security parameter $s$. However, the message generated by the user is expressed by,

$$\widetilde{M} = h(U_{ID} \| s) \oplus U_{ps} \qquad (1)$$

The user ID and system security parameters are concatenated and operated with the hash function. Then, the password is applied to perform the EX-OR function. Then, the resultant hashing and EXOR function is performed to determine $\widetilde{M}$. The message received by the user is given by,

$$M = h(U_{ID} \| s) \oplus U_{ps} \qquad (2)$$

where the term $h$ refers to the hashing function and the user ID is denoted as $U_{ID}$ the user password is indicated as $U_{ps}$, and the symbol $\oplus$ refers to Ex-OR operation. If $M = \widetilde{M}$ then, the registration request sends by the user to the cloud server is accepted in such a way that the user is registered with a cloud server. Then, the system generates the parameter securely and save $s$ and $m$. Once the setup phase is completed, the key is generated.

*3.2 Key generation phase*

The second phase of the proposed model is the key generation phase. Here, the KGC generates the s public key of the user $U^P$ with the user attribute $a$. Here, the user attribute value ranging from $0 \text{ and } 1$, which means $a \in [0,1]$. Besides, the KGC produced the private key of the user using the public key, system security parameter, and the user attribute. Thus, the generated private key is forwarded to the cloud server and user and save it $U_{pk}^*$. However, the private key of the user, generated by the KGC is expressed as,

$$U_{pk} = m \bmod (U^P \oplus (s \| a)) \qquad (3)$$

The system security parameter that is stored in the KGC is then concatenated with the user attribute and is allowed to perform Ex-OR operation with the public key of the user. Then, the master key $m$ is modulated with the obtained output to form the private key of the user.

*3.3 Encryption phase*

The next phase in the proposed access control-based privacy preservation method is the encryption phase. Here, the entities, such as the data owner and cloud server, are used in the encryption phase. Here, the term $D$ is considered as the input data, which is encrypted in the encrypted phase by the data owner. The data $D$ encrypted by the data owner is denoted as $D^E$, and the equation is given below,

$$D^E = E(D) \oplus h(s \| m) \qquad (4)$$

The input data $D$ is allowed to perform encryption function, and the security parameter is concatenated with the master key, and the resulted expression is applied to the hashing function. The output obtained from the hashing function and encrypted data is allowed for performing Ex-OR operation in order to encrypt data. Moreover, the data owner produces the cipher-text $D^C$ based on encrypted data and the private key of the user $O_{pk}$ of the data owner. The generated cipher-text by data owner is given by,

$$D^C = E(D^E \| O_{pk}) \quad (5)$$

The encrypted data is concatenated with the private key of the user and is given to the encryption function to generate the cipher-text $D^C$. Once the encrypted data is identified by the data owner and is forwarded to the cloud server, so where it is stored as the $D^{C^*}$. After that, the user begins to execute the access control phase.

*3.4 Access control phase*

In this phase, the access control is run by the three entities, such as user, cloud server, and KGC. Here, the data access query $q$ is determined through the hashing, and the concatenation operation were three parameters, like $\tilde{M}$, $U_{ID}$ and $U_{pk}^*$. Hence, it is done based on the following equation to generate $q$, and is expressed as,

$$q = \tilde{M} * h(U_{ID} \| U_{pk}^*) \quad (6)$$

Once the data access query $q$ is identified based on the following equation that is forwarded to the cloud server. Here, the cloud server tries to point out $\tilde{q}$, and the equation is given below,

$$\tilde{q} = M * h(U_{ID}^* \| U_{pk}^*) \quad (7)$$

The user ID of the cloud server is concatenated with the private key of the cloud server and is then forwarded to the hashing function. The output obtained from the hashing function is then applied to $M$ in order to generate $\tilde{q}$. If $q = \tilde{q}$ from the cloud server, the user request is accepted, then the session key of the user $U_{sk}$ is generated in the KGC entity. To generate $U_{sk}$, the master key, user attribute, and the public key of the user is required, and the equation is given below,

$$U_{sk} = E(U^P \| h(m \| a)) \quad (8)$$

The master key $m$ is concatenated with the user attribute $a$ and is fed to the hashing function. The resultant expression generated by the hashing function is concatenated with the public key of the user and is applied to the encrypted function in order to generate the session key of the user $U_{sk}$. Hence, the generated $U_{sk}$ is subjected to the cloud server and user, which is saved $U_{sk}^*$. After performing the access control phase, the next step to be performed is the validation phase.

*3.5 Validation phase*

In the validation phase, the validation messages $v_1$ and $v_2$ are determined. Hence, through the parameters user ID $U_{ID}$, session key $U_{sk}$, and system security parameter $s$. Similarly, for $v_2$ the message, the validation message $v_2$ is followed based on user ID, saved the private key of the user, and the master key. Thus, the expression of the validation messages is given by,

$$v_1 = h(U_{ID} \| U_{sk} \| s) \bmod r \quad (9)$$
$$v_2 = h(U_{ID} \| U_{pk}^* \| m) \bmod a \quad (10)$$

To identify the validation message $v_1$, the user ID $U_{ID}$, session key $U_{sk}$, and system security parameters $s$ are concatenated and are applied to the hashing function. The output obtained from the hashing function is then modulated with the random number where the value of random number ranging from $0$ to $1$, whereas for $v_2$ user ID, saved private key of a user $U_{pk}^*$, and the master key is concatenated, and hashing function is applied. Then the resultant expression is modulated with the master key. After that, $v_1$ and $v_2$ are transformed into the cloud server to find out $\tilde{v}_1$ and $\tilde{v}_2$. The $\tilde{v}_1$ and $\tilde{v}_2$ are identified based on the parameter, which is already received from the private channel of the user, and the equation is given by,

$$\tilde{v}_1 = h(U_{ID} \| U_{sk} \| s) \bmod r \quad (11)$$
$$\tilde{v}_2 = h(U_{ID} \| U_{pk}^* \| m) \bmod a \quad (12)$$

Once $\tilde{v}_1$ and $\tilde{v}_2$ are determined through the private channel, the matching is performed as $v_1 = \tilde{v}_1$, and $v_2 = \tilde{v}_2$. If both conditions are satisfied, the user is verified. Once the verification phase is carried out, the user begins to access the data and data sharing phase.

*3.6 User data access and data sharing phase*

Once the user is verified, the data is shared with the user. Here, the two entities, such as the cloud server and the user, are considered. The $D^{E^*}$ is determined in the user based on the decryption of $D^C$, and the equation is given by,

$$D^{E^*} = DE(D^C) \quad (13)$$

Then, the $D^*$ is identified based on the following equation,

$$D^* = DE(D^{E^*} \oplus h(s \| m)) \quad (14)$$

The system security parameter and the master key are concatenated with the hashing function, and then the Ex-

OR function is applied to $D^{E^*}$. The output expression is decrypted to form $D^*$. Table 1 illustrates the proposed access control flow for privacy preservation in the cloud infrastructure.

## 4. RESULTS AND DISCUSSION

The result of the proposed access control-based privacy preservation method is discussed with respect to memory usage and a genuine detection rate.

### 4.1 Experimental setup

The implementation of the developed model is carried out in the Python tool with windows 10 OS, 4GB RAM, and Intel I3 processor.

### 4.2 Dataset description

The experiment was performed based on the Swiss, Hungarian, and Cleveland dataset were taken from Heart disease dataset [21]. The Cleveland database was taken from the Cleveland Clinical Foundation contributed by David W. Aha. The Hungarian dataset was obtained from the Hungarian Institute of Cardiology. The Swiss dataset was obtained from the university hospital, Basel, Switzerland. The dataset comprises of 303 instances with 76 attributes. Among the 76 attributes available, 14 of them are used for most of the researches.

### 4.3 Performance metrics

The performance of the proposed access control-based privacy preservation method is employed for analyzing the methods includes memory usage and genuine detection rate.

*4.3.1. Memory usage:* It is the system memory, which is utilized to measure the usage of memory for recording and to process the data.

*4.3.2. Genuine detection rate:* It is utilized to measure the number of genuine users as compared to the total number of users.

### 4.4 Comparative Methods

The methods employed for the analysis include: Privacy preserving-based healthcare storage system [1], SEEMAKA [3], H-CLSC [6], and proposed access control-based privacy preservation method.

*a) Comparative analysis based on Hungarian dataset*

Figure 2 portrays the comparative analysis of the proposed access control-based privacy preservation method using the Hungarian dataset in terms of metrics. Figure 2(a) depicts the comparative analysis of memory usage by varying the key length. When the key length=64, the memory usage obtained by the existing privacy preserving-based healthcare storage system is 0.0639GB, SEEMAKA is 0.0638GB, H-CLSC is 0.06387GB, while the proposed access control-based privacy preservation method obtained lower memory usage of 0.06386GB, respectively. Figure 2(b) depicts the comparative analysis of the Genuine detection rate with respect to the key length.

When key length=64, the Genuine detection rate obtained by the existing privacy preserving-based healthcare storage system is 0.73, SEEMAKA is 0.82, H-CLSC is 0.85, while the proposed access control-based privacy preservation method obtained higher Genuine detection rate of 0.91, respectively.

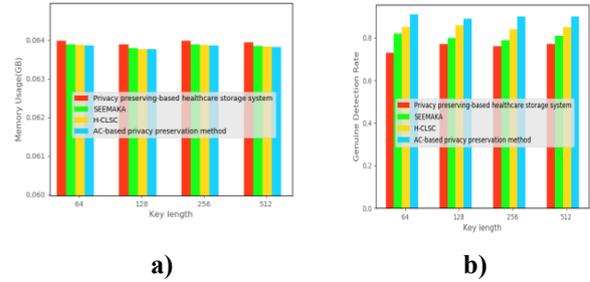

**Figure 2.** Analysis of methods by varying the key length based on Hungarian dataset a) memory usage, and b) Genuine detection rate

*b) Comparative analysis based on Cleveland dataset*

Figure 3 presents the analysis of methods by varying the key length in terms of performance metrics using the Cleveland dataset. Figure 3a) depicts the comparative analysis of memory usage by varying the key length. When key length=512, the memory usage of methods computed by privacy preserving-based healthcare storage system, SEEMAKA, H-CLSC, and proposed access control-based privacy preservation method are 0.06428GB, 0.06424GB, 0.06422GB, and 0.06421GB. Figure 3b) depicts the comparative analysis of the Genuine detection rate with respect to the key length. When key length=512, the Genuine detection rate obtained by the existing privacy preserving-based healthcare storage system is 0.75, SEEMAKA is 0.81, H-CLSC is 0.83, while the proposed access control-based privacy preservation method obtained higher Genuine detection rate of 0.89, respectively.

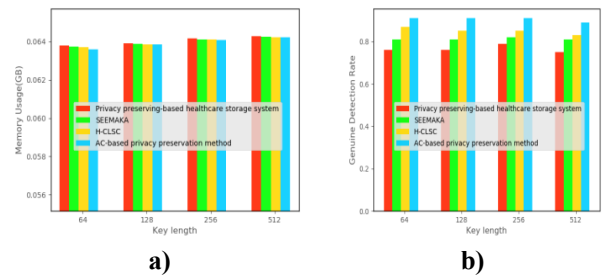

**Figure 3.** Analysis of methods by varying the key length based on Cleveland dataset a) memory usage, and b) Genuine detection rate

*c) Comparative analysis based on Swiss dataset*

Figure 4 depicts the comparative analysis in terms of memory usage and the Genuine detection rate using the Swiss dataset. Figure 4a) depicts the comparative analysis of memory usage by varying the key length. When key

length=512, the memory usage of methods computed by privacy preserving-based healthcare storage system, SEEMAKA, H-CLSC, and proposed access control-based privacy preservation method are 0.0639GB, 0.06388GB, 0.06387GB, and 0.06385GB. Figure 4b) depicts the comparative analysis of the Genuine detection rate with respect to the key length. When key length=512, the Genuine detection rate obtained by the existing privacy preserving-based healthcare storage system is 0.79, SEEMAKA is 0.81, H-CLSC is 0.84, while the proposed access control-based privacy preservation method obtained higher Genuine detection rate of 0.89, respectively.

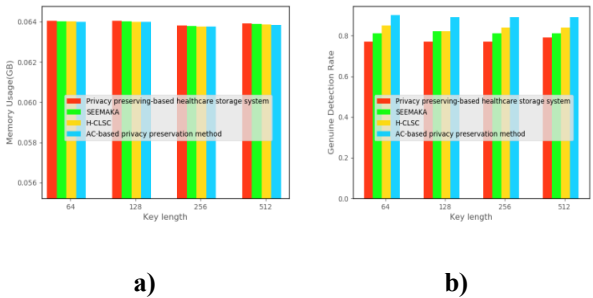

a)          b)

**Figure 4.** Analysis of methods by varying the key length based on Swiss dataset a) memory usage, and b) Genuine detection rate

*4.5. Comparative discussion*

Table 2 illustrates a comparative discussion using memory usage and a genuine detection rate parameter by varying the key length. The minimal memory usage of 0.0636GB obtained by the proposed access control-based privacy preservation method, whereas the existing privacy preserving-based healthcare storage system, SEEMAKA, and H-CLSC are 0.06378GB, 0.06374GB, and 0.06372GB based on Cleveland dataset. The performance of the proposed access control-based privacy preservation method, the maximal genuine detection rate of 0.91, is obtained by the Hungarian dataset.

5. CONCLUSION

In this study, a robust and effective privacy preservation approach, namely "access control-based privacy preservation model," has been proposed to improve data sharing in cloud infrastructure. The developed model involves six phases, such as setup, key generation, encryption, validation, access control, and data sharing. Also, the proposed access control-based privacy preservation model consists of four different entities, such as user, data owner, cloud server, and KGC. Each entity in data sharing performs its operations for sharing the data in the cloud by maintaining privacy.

The user, such as healthcare staff for the patient's friends, relatives registers to the medical institute to obtain secret keys. The cloud server is responsible for storing healthcare big data for numerous medical institutes and replies to the data access queries. The developed model attained minimal memory usage of 0.0636GB based on Cleveland dataset, and maximal genuine detection rate of 0.91 by considering Hungarian dataset. In future work, additional entities in the cloud will be included to explore the efficiency of existing methods further.

| Datasets | Metrics | Health-care storage system | SEEMAKA | H-CLSC | Proposed |
|---|---|---|---|---|---|
| Hungarian | *Memory usage (GB)* | 0.0639 | 0.0638 | 0.06387 | **0.06386** |
| Hungarian | *Genuine detection rate* | 0.73 | 0.82 | 0.85 | **0.91** |
| Cleveland | *Memory usage (GB)* | 0.06378 | 0.06374 | 0.06372 | **0.0636** |
| Cleveland | *Genuine detection rate* | 0.76 | 0.81 | 0.87 | **0.91** |
| Swiss | *Memory usage (GB)* | 0.06405 | 0.06402 | 0.06401 | **0.0639** |
| Swiss | *Genuine detection rate* | 0.77 | 0.81 | 0.85 | **0.9** |

Table 2. Comparative analysis of memory usage and genuine detection.

| Phases | User | Data Owner | Cloud server | KGC |
|---|---|---|---|---|
| **Setup phase** | $U_{ID}, U_{ps}$ <br> $\tilde{M} = h(U_{ID} \| s) \oplus U_{ps}$ | $\tilde{M}$ | $U^*_{ID}, U^*_{ps}$ <br> $M = h(U_{ID} \| s) \oplus U_{ps}$ <br> If $M = \tilde{M}$, <br> User is registered | Generate system security parameter $(s)$ and $m$. |
| **Key generation** | | | $U^*_{pk}$ | Generates $U^P$ and $a \in [0,1]$ <br> $U_{pk} = m \mod \begin{pmatrix} U^P \oplus \\ (s\|a) \end{pmatrix}$ |
| **Encryption** | $U^*_{pk}$ | $D^E = E(D) \oplus h(s\|m)$ <br> $D^C = E(D^E \| O_{pk})$ | $D^{C*}$ | |
| **Access control** | $q = \tilde{M} * h(U_{ID} \| U^*_{pk})$ <br> $U^*_{sk}$ | | $\tilde{q} = M * h(U^*_{ID} \| U^*_{pk})$ <br> If $q = \tilde{q}$, <br> User request is accepted. <br> $D^{C*}$ | Generates <br> $U_{sk} = E\begin{pmatrix} U^P \| h \\ (m\|a) \end{pmatrix}$ |
| **Validation** | $v_1 = h(U_{ID} \| U_{sk} \| s) \mod r$ <br> $v_2 = h(U_{ID} \| U^*_{pk} \| m) \mod a$ | | $v_1, v_2$ <br> $\tilde{v}_1 = h(U_{ID} \| U_{sk} \| s) \mod r$ <br> $\tilde{v}_2 = h(U_{ID} \| U^*_{pk} \| m) \mod a$ <br> If $v_1 = \tilde{v}_2$ and $v_2 = \tilde{v}_2$ <br> User is verified | |
| **Data sharing** | User access the data and decrypts. <br> $D^{E*} = DE(D^C)$ <br> $D^* = DE(D^{E*} \oplus h(s\|m))$ | | Once user is verified, the data is shared to the user. <br> $D^C$ | |

**Table1.** Proposed access control flow for privacy preservation in cloud